\documentclass[%
reprint,
superscriptaddress,
amsmath,amssymb,
aps,
pra,
]{revtex4-1}

\usepackage{amsfonts}
\usepackage{mathrsfs}
\usepackage{amsmath}
\usepackage{color}
\usepackage{graphicx}
\usepackage{bm}
\usepackage{amssymb}
\usepackage{xspace}
\usepackage{epstopdf}
\usepackage{dcolumn}
\usepackage{longtable}
\usepackage{ulem} 
\usepackage{multirow}
\usepackage[colorlinks=true, pdfstartview=FitV, linkcolor=red, citecolor=blue, urlcolor=blue]{hyperref}
\usepackage[colorinlistoftodos,prependcaption,textsize=tiny]{todonotes}

\newcommand{\beq}{\begin{equation}}
\newcommand{\eeq}{\end{equation}}
\def\bse{ \begin{subequations} }
\def\ese{ \end{subequations} }
\def\beqa{ \begin{eqnarray} }
\def\eeqa{ \end{eqnarray} }

\begin{document}

\title{Bright solitons in a spin-tensor-momentum-coupled Bose-Einstein condensate}

\author{Jie Sun}
\affiliation{International Center of Quantum Artificial Intelligence for Science and Technology (QuArtist)
and Department of Physics, Shanghai University, Shanghai 200444, China}

\author{Yuanyuan Chen}
\email{cyyuan@staff.shu.edu.cn}
\affiliation{International Center of Quantum Artificial Intelligence for Science and Technology (QuArtist)
and Department of Physics, Shanghai University, Shanghai 200444, China}

\author{Xi Chen}
\email{xchen@shu.edu.cn}

\affiliation{International Center of Quantum Artificial Intelligence for Science and Technology (QuArtist)
	and Department of Physics, Shanghai University, Shanghai 200444, China}

\affiliation{Department of Physical Chemistry, University of the Basque Country UPV/EHU, Apartado 644, 48080 Bilbao, Spain}

\author{Yongping Zhang}
\email{yongping11@t.shu.edu.cn}
\affiliation{International Center of Quantum Artificial Intelligence for Science and Technology (QuArtist)
and Department of Physics, Shanghai University, Shanghai 200444, China}

\begin{abstract}
Synthetic spin-tensor-momentum coupling has recently been proposed to realize in atomic Bose-Einstein condensates. Here we study bright solitons in Bose-Einstein condensates with spin-tensor-momentum coupling and spin-orbit coupling. The properties and dynamics of spin-tensor-momentum-coupled and spin-orbit-coupled bright solitons are identified to be different. We contribute the difference to the different symmetries.
\end{abstract}

\maketitle

\section{Introduction}

In ultracold neutral atoms, hyperfine spin states,
coupling to linear momentum ~\cite{Spielman,Pan,Wang,Cheuk,Engels,Olson,Hamner} or orbital angular momentum~\cite{Lin, JiangK},
are interesting and significant not only in
fundamental phenomena of ultracold atoms and condensed matter physics, but also in the applications in quantum information processing, atom metrology and atomitronics, with the current
experimental progress. Particularly, the spin-orbit coupling (SOC)
provides the unique dispersion relationship, exhibiting particular features without analogues in the case of without the SOC. The competition between atomic many-body interactions and the dispersion relation generates many fundamental ground state phases~\cite{Zhai,Ho,Wucongjun,Sinha,Yongping1,Hu, Li2012,Ketterle} and exotic collective excitations~\cite{Khamehchi,Pan2015} in spin-orbit-coupled Bose-Einstein condensates (BECs).

The interplay between the nonlinearity stemming from atomic interactions and dispersions also gives rise to the existence of bright solitons which are spatially localized states. The interested spin-orbit-coupled dispersions inevitably change the existence and properties of bright solitons~\cite{Achilleos,Xuyong1}. In general, solitons follow the symmetries of spin-orbit-coupled Hamiltonian, which provides a deep insight into the searching  of solitons. Moreover, the dynamics of solitons is always accompanied by rich spin dynamics~\cite{Fialko,Wen}. The lack of Galilean invariance in spin-orbit-coupled systems~\cite{BiaoWu} makes that it is nontrivial to find movable solitons, one can not directly obtain a movable soliton from its stationary correspondence. Different aspects of bright solitons with the SOC have been investigated a lot~\cite{Salasnich,Zhou,Gautam,Sakaguchi,Sakaguchi2,Kartashov}, ranging from with long-ranged dipole interactions~\cite{Xuyong2,Liyongyao,Chiquillo2} to in optical lattices~\cite{Kartashov2,Lobanov,Busch,Sakaguchi3,LiHong,Mardonov,Kartashov3}.

Very recently, the generation of  artificial spin-tensor-momentum coupling (STMC) into an atomic BEC has been proposed~\cite{Luo}. Different from the usual spin-orbit coupling where linear momentum is coupled with spin vectors, STMC is the interaction between linear momentum and spin tensors. Such emergent interaction can be applicable to the discovery of exotic topological matters~\cite{HaipingHu,Chengang}.

In this paper, we investigate bright solitons in STMC BECs in which the three components of the ground hypefine states of $^{87}$Rb are utilized for experimental implementation.
We first apply imaginary-time evolution method to
study the  stationary properties of STMC soliton,
and further explore the dynamics by using variational method.
By comparing with SOC bright soliton in Refs.~\cite{Liu,Gautam2,Xue},  we
conclude that the difference between STMC and SOC bright solitons originates from the different symmetry relevant to spin rotation.

The paper will be organized as follows. In Sec. \ref{model} the systems and Hamiltonian are introduced for SOC and STMC. Here we present both for completeness and further comparison. Later, the bright solitons are discussed for both STMC and SOC BECs in Sec. \ref{soliton}, to clarify the difference in the spin rotation and symmetry. Finally, conclusion are
made in Sec. \ref{conclusion}.

\section{Model and Hamiltonian}
\label{model}

\begin{figure}[t]
\includegraphics[bb=0 13 563 460, width=0.45\textwidth]{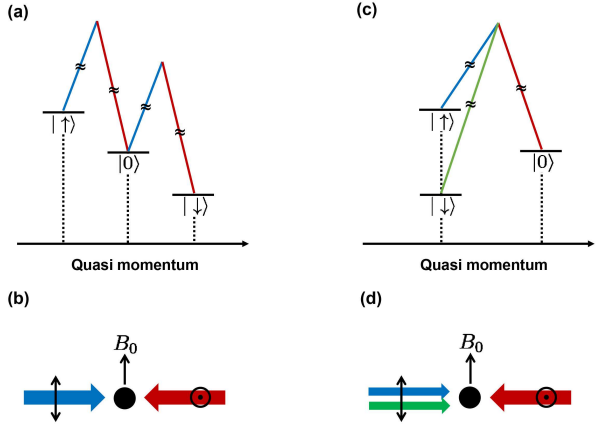}
\caption{Experimental schemes to realize the spin-orbit coupling (a,b) and spin-tensor-momentum coupling (c,d). Three hyperfine states ($|\uparrow \rangle,   | 0 \rangle, |\downarrow \rangle $) are split by a bias magnetic field $B_0$. In (a,b) two laser beams propagate oppositely to couple $|p_x-2\hbar k_\text{R}, \uparrow \rangle,   |p_x,  0 \rangle, |p_x+2\hbar k_\text{R}, \downarrow \rangle $ with $p_x$ being momentum along laser direction and quasimomentum $2\hbar k_\text{R}$ relevant to the wavenumber of lasers, the quasimoentum difference between hyperfine states constitutes the spin-orbit coupling. In (c,d) two beams whoes polarizations are parallel to the bias magnetic field propagate along same direction and the third beams in the opposite direction. They can couple $|p_x-2\hbar k_\text{R}, \uparrow \rangle,   |p_x,  0 \rangle, |p_x-2\hbar k_\text{R}, \downarrow \rangle $. }
\label{Scheme}
\end{figure}

We first consider the experiment of synthetic SOC in three-component BECs~\cite{Spielman2,Spielman3}, where the three hyperfine states of $^{87}$Rb atoms are utilized, with the energy splitting by a bias magnetic field, as shown in Fig.~\ref{Scheme}(a,b). To realize SOC, the atoms are dressed by two counterpropagating Raman laser beams, and the polarizations of lasers are arranged so that  two-photon optical transitions can
be induced, see Fig.~\ref{Scheme}(b).
The transitions in the basis of $( | \uparrow \rangle=|1,-1\rangle,   | 0 \rangle=|1,0\rangle, | \downarrow \rangle=|1,1\rangle  )$ are engineered as,
\begin{equation}H_\text{Ram}^\text{SOC}=\Omega
\begin{pmatrix}
0 & e^{-i2k_\text{R}x}&0\\  e^{i2k_\text{R}x}&0& e^{-i2k_\text{R}x} \\ 0&  e^{i2k_\text{R}x}&0
\end{pmatrix}, \notag
\end{equation}
where $\Omega$ is the strength of two-photon Rabi coupling~\cite{Yongping2} and $k_\text{R}$ is the wavenumber of the Raman beams. During the transitions, there is a momentum exchange between the atoms and lasers. Including kinetic energy, the Hamiltonian becomes,
$
H_\text{SOC}=p_x^2/2m +  H_\text{Ram}^\text{SOC},
$,
with $m$ being atomic mass and $p_x$ being momentum along the direction of Raman lasers. To explicitly show the existence of SOC, a unitary transformation is needed, $U_\text{SOC}=e^{i2 k_\text{R}x F_z} $, such that the Hamiltonian $\tilde{H}_\text{SOC} =U_\text{SOC} H_\text{SOC} U_\text{SOC}^{-1} $ becomes
\beq
\label{SOC}
\tilde{H}_\text{SOC} =\frac{p_x^2}{2m} -\frac{4\hbar k_\text{R} p_xF_z }{2m} +\frac{4(\hbar k_\text{R})^2F_z^2 }{2m}+ \sqrt{2} \Omega F_x.
\eeq
Here $(F_x, F_y, F_z)$ are spin-1 Pauli matrices, and the SOC $2\hbar k_\text{R} p_xF_z/m$ is involved. Physically, the SOC means that there is a quasimomentum difference $-2\hbar k_\text{R}$  between states  $| \uparrow \rangle$ and  $  | 0 \rangle$, and between  $  | 0 \rangle$ and $ | \downarrow \rangle  $.

Next, the STMC can be introduced artificially by dressing the atoms with three Raman beams~\cite{Luo}, see Fig.~\ref{Scheme}(c,d). Two of them with same linear polarization propagate along same direction,  and the other propagates oppositely. The two-photon transitions accompanying momentum transfers become,
\begin{equation}H_\text{Ram}^\text{STMC}=\Omega
\begin{pmatrix}
0 & e^{-i2k_\text{R}x}&0\\  e^{i2k_\text{R}x}&0& e^{i2k_\text{R}x} \\ 0&  e^{-i2k_\text{R}x}&0
\end{pmatrix}. \notag
\end{equation}
Note that the difference between $H_\text{Ram}^\text{SOC}$ and $H_\text{Ram}^\text{STMC}$ is very slight. To eliminate the spatial dependence in  $H_\text{Ram}^\text{STMC}$, a unitary transformation $U_\text{STMC}=e^{i2 k_\text{R}x F_z^2} $ is performed, and the new total Hamiltonian $ \tilde{ H}_\text{STMC}= U_\text{STMC} H_\text{STMC}U_\text{STMC}^{-1}$, with
$  H_\text{STMC}=p_x^2/2m +  H_\text{Ram}^\text{STMC} $
is expressed as,
\beq
\label{STMC}
\tilde{ H}_\text{STMC}=\frac{p_x^2}{2m} -\frac{4\hbar k_\text{R} p_xF_z^2 }{2m} +\frac{4(\hbar k_\text{R})^2F_z^2 }{2m}+ \sqrt{2} \Omega F_x. 
\eeq
The STMC takes a specific form as $2\hbar k_\text{R} p_xF_z^2/m$. From the above equation, it is clear that such specific STMC is just a rearrangement of quasimomentum difference comparing with the case of the SOC. The quasimomentum difference between $| \uparrow \rangle$ and  $  | 0 \rangle$ is $-2\hbar k_\text{R}$, while it is  $2\hbar k_\text{R}$  between  $  | 0 \rangle$ and $ | \downarrow \rangle$.

\section{Bright Solitons with STMC and SOC}

\label{soliton}

Now, we are ready to study bright solitons in the BECs with both the STMC and SOC whose experimental realizations are analyzed in the previous section \ref{model}. We start from the standard Gross-Pitaevskii (GP) equations and take into consideration the spin-tensor-momentum-coupled and spin-orbit-coupled Hamiltonian in Eq.~(\ref{SOC}) and Eq.~(\ref{STMC}).
The dimensionless GP equations for spin-tensor-momentum-coupled BEC are,
\begin{equation}
\label{GPSTMC}
i\frac{\partial \Psi}{\partial t}= [-\partial_{x}^{2}+(4i\partial_{x}+4 +\Delta)F_{z}^{2}+\sqrt{2}\Omega F_{x}+H_\text{int}]\Psi,
\end{equation}
while, the spin-orbit-coupled GP equations are
\begin{equation}
\label{GPSOC}
i\frac{\partial \Psi}{\partial t}= [-\partial_{x}^{2}+4i\partial_{x}F_{z}+(4+\Delta)F_{z}^{2}+\sqrt{2}\Omega F_{x}+H_\text{int}]\Psi.
\end{equation}
In the both equations, the units of energy, position coordinate and time that we adopt are $   \hbar^2 k_\text{R}^2/2m, 1/k_\text{R}  $ and $   2m /\hbar k_\text{R}^2  $ respectively. The additional term $\Delta F_{z}^{2}$ originates from quadratic Zeeman effect. Three-component wave functions are $ \Psi=( \Psi_{\uparrow},\Psi_{0},\Psi_{\downarrow}   )^T$, for convenience, in the following, we relabel the wave functions as  $ \Psi=( \Psi_1,\Psi_2,\Psi_3   )^T$. In above equations, $H_\text{int}=g_0( |\Psi_1|^2 +|\Psi_2|^2+|\Psi_3|^2 )$, for simplicity, we only consider interactions having SU(3) symmetry. Since our aim is to investigate the bright solitons, we focus on attractive interactions of $g_0<0$.

\begin{figure*}[t]
\includegraphics[width=0.6\textwidth]{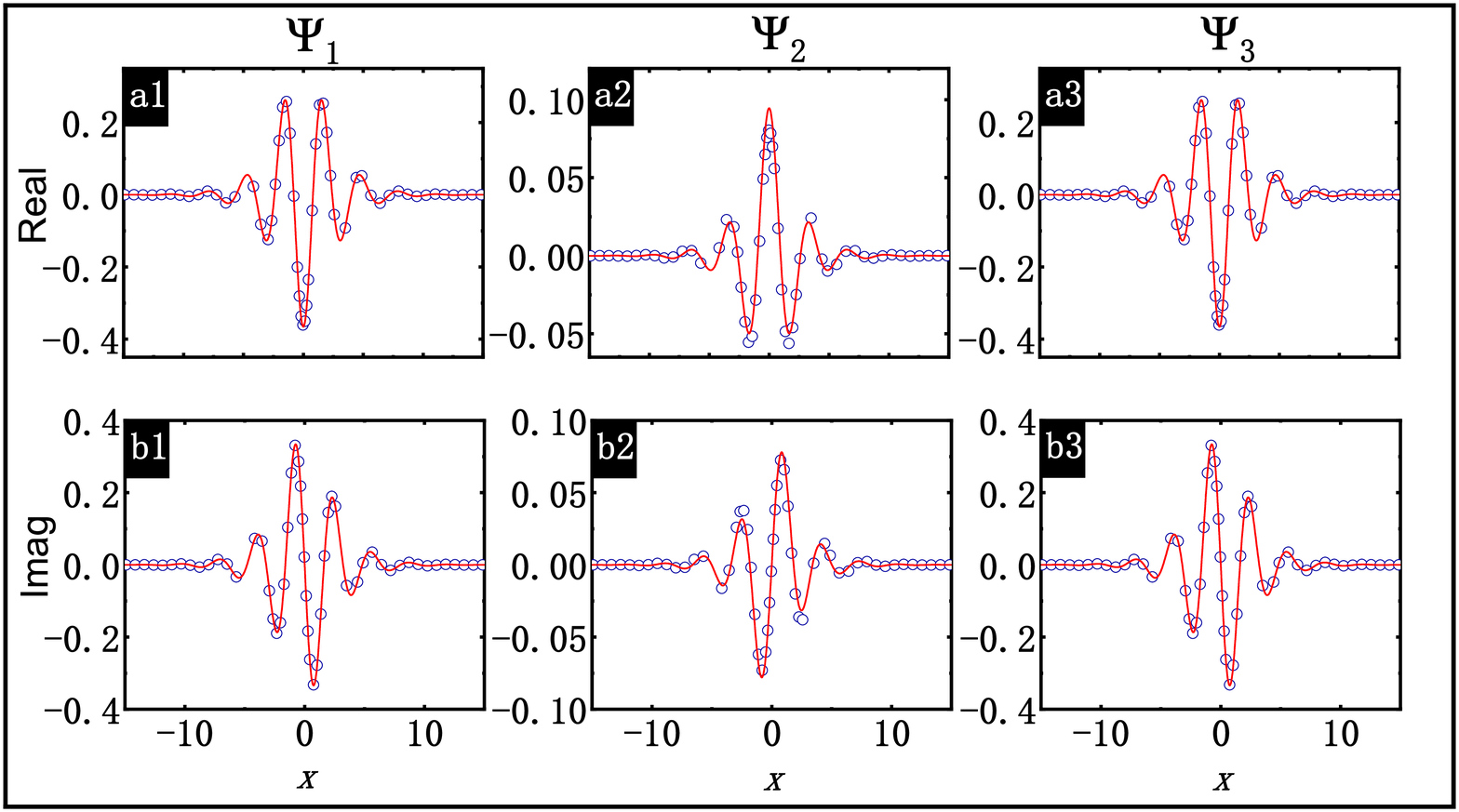}
\includegraphics[width=0.6\textwidth]{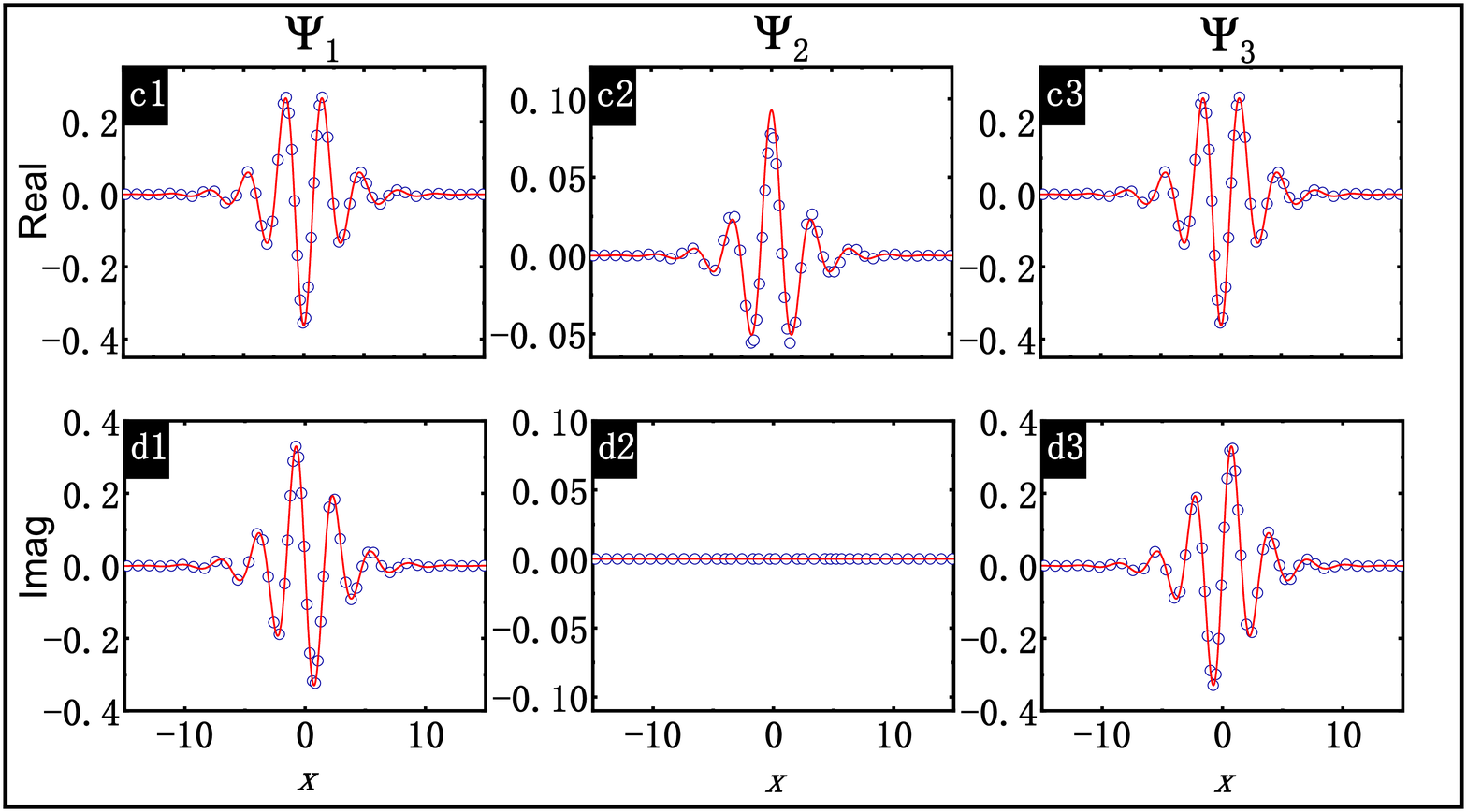}
\caption{Profiles of the spin-tensor-momentum-coupled (upper panel) and spin-orbit-coupled (lower panel) bright solitons. In each panel, the first (second) row is the real (imaginary) parts of soliton wave functions $ \Psi=( \Psi_1,\Psi_2,\Psi_3)^T$. Solid-lines are solutions from the imaginary-time evolution method and dot-lines are analytical solutions from the variational method.  The dimensionless parameters are $\Delta=-1, \Omega=0.5$ and $g_0=-2$.}
\label{Soliton}
\end{figure*}

The difference between the spin-tensor-momentum-coupled and spin-orbit-coupled GP equations is the appearance of $4i\partial_{x}F_{z}^{2}$ and $4i\partial_{x}F_{z}$. Such difference leads to different symmetries of GP equations, which affects the properties of bright solitons. We find stationary bright solitons by the numerical calculation of GP equations using the imaginary-time evolution method, because of which, the soliton solutions belong to ground states. Typical soliton profiles are demonstrated in Fig.~\ref{Soliton}. The upper panel is the profiles of spin-tensor-momentum-coupled solitons, and the lower panel is that of spin-orbit-coupled solitons. For further comparison, we adopt same parameters for the GP equations with STMC and SOC. Our general observation is that the imaginary parts of soliton wave functions for both STMC and SOC do not vanish. In contrast, the ground states of ordinary BECs (without STMC or SOC) are real-valued with no node in wave functions~\cite{Wucongjun}. This is the unique feature of spin-orbit-coupled~\cite{Xuyong1} and spin-tensor-momentum-coupled BECs. At first sight, the spin-tensor-momentum-coupled solitons share same profiles with spin-orbit-coupled solitons, especially, the real parts of soliton wave functions are almost same. However, there exists an apparent difference in the imaginary parts.

Our solitons as ground states follow symmetries of the systems. The stationary spin-tensor-momentum-coupled GP equations in Eq.~(\ref{GPSTMC}) have a spin rotating symmetry,
\begin{equation}
\mathcal{R}_\text{STMC}=e^{i\pi F_x}=\begin{pmatrix} 0 & 0&-1 \\ 0&-1&0 \\-1 &0&0 \end{pmatrix},
\end{equation}
which rotates spins along the $F_x$ axis by the angle of $\pi$, and a joint parity symmetry,
\begin{equation}
\mathcal{O}_\text{STMC}=\mathcal{P}\mathcal{K},
\end{equation}
with $\mathcal{P}$ and $\mathcal{K}$ being the parity and complex conjugate operators. The symmetry $\mathcal{R}_\text{STMC}$ is relevant to the spin tensor $F_x^2$, since $F_x^2=\frac{1}{2}(\mathbb{I}- \mathcal{R}_\text{STMC})$. The eigen-equation is $\mathcal{R}_\text{STMC} \Psi= \pm \Psi$. For the $+1$ eigenstate, $\Psi_2(x)=0$, which leads to $\langle F_x \rangle =0$. Whereas, to minimize energy of Rabi coupling term $\sqrt{2} \Omega F_x$, it is preferable that $\langle F_x \rangle <0$. Therefore, bright solitons select the eigenstate with $-1$ eigenvalue, $\mathcal{R}_\text{STMC} \Psi=  -\Psi$, the consequence of which is $\Psi_1(x)=\Psi_3(x)$. Fig.~\ref{Soliton} demonstrates $\Psi_1(x)=\Psi_3(x)$ from the real and imaginary parts. The symmetry $\mathcal{O}_\text{STMC}$ determines that the parity of real parts of soliton wave functions $\Psi_1,\Psi_2$ and $\Psi_3$ should be opposite to that of imaginary parts. The real parts are even and imaginary parts are odd, see Fig.~\ref{Soliton}.

\begin{figure*}[t]
\includegraphics[width=0.76\textwidth]{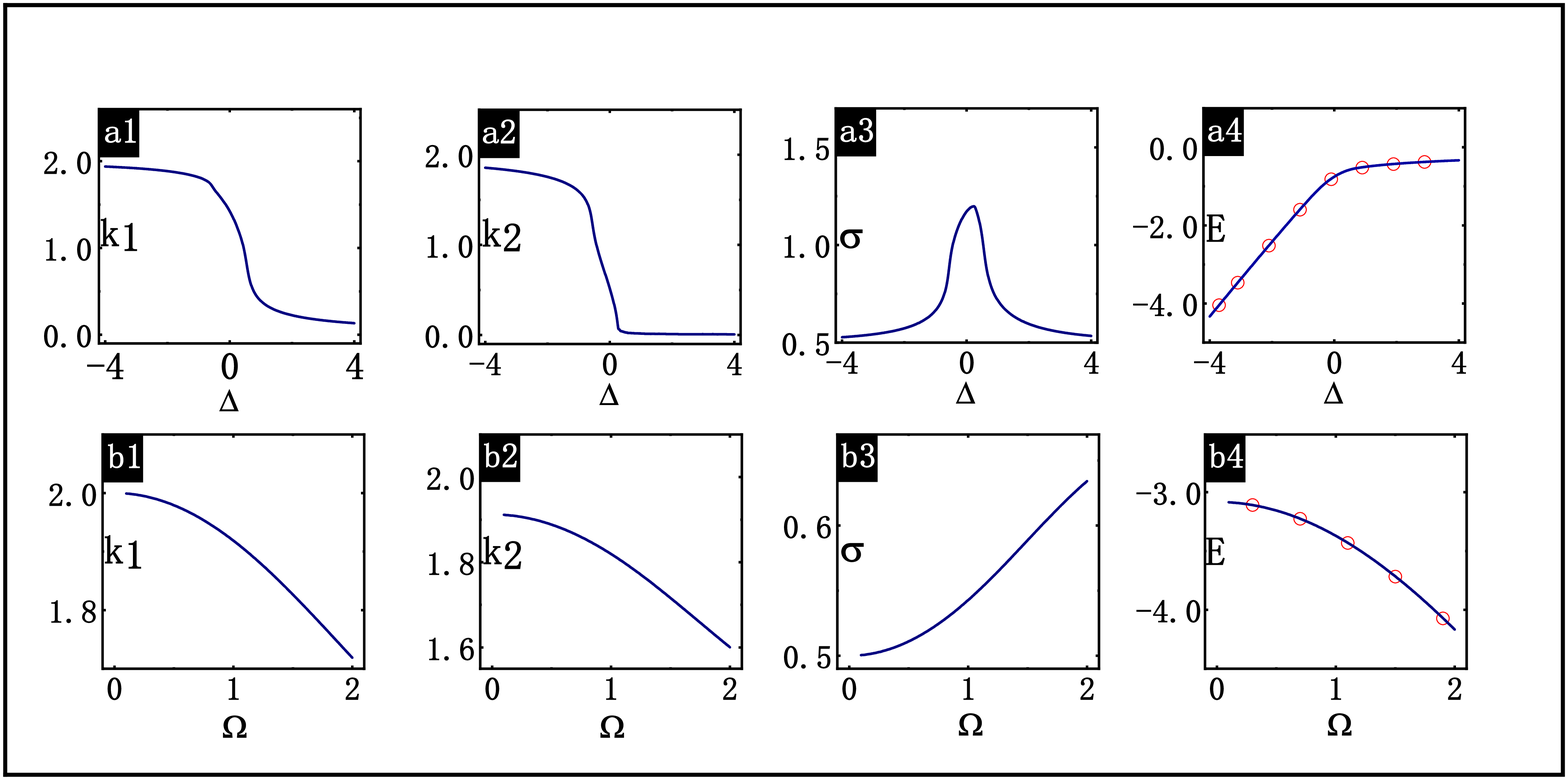}
\includegraphics[width=0.76\textwidth]{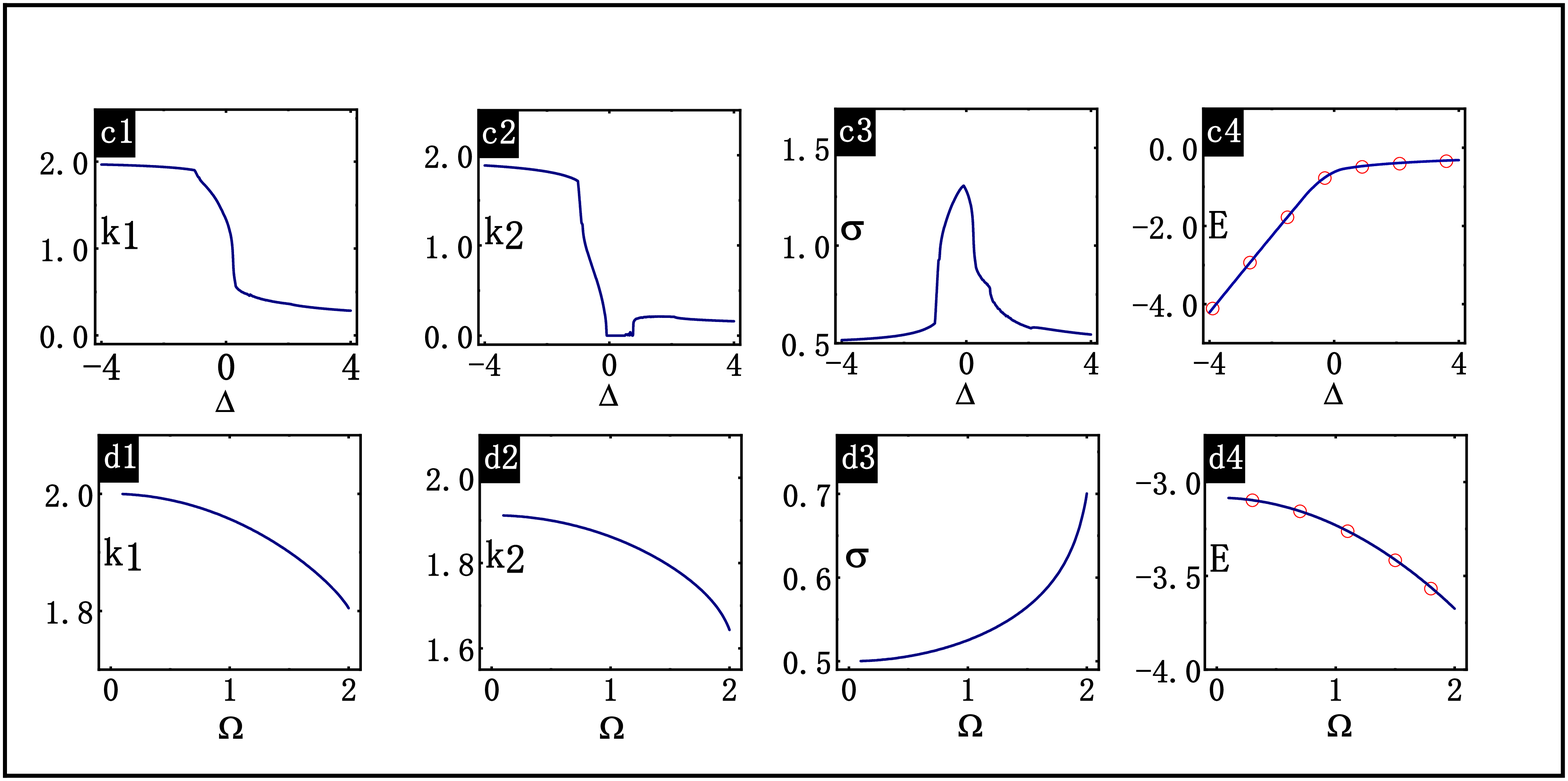}
\caption{Features of the spin-tensor-momentum-coupled (upper panel) and spin-orbit-coupled (lower panel) bright solitons characterized from variational wave functions. The variational parameters $k_1,k_2,\sigma$ and total energy $E_\text{STMC}, E_\text{SOC}$ are a function of $\Delta$ and $\Omega$. Solid-lines are from the variational method and dots are from the imaginary-time evolution method. In the first (second) row of each panel, $\Omega=1$ ($\Delta=-3$). The nonlinear coefficient $g_0=-2$. }
\label{Feature}
\end{figure*}

The symmetry of the stationary spin-orbit-coupled GP equations in Eq.~(\ref{GPSOC}) is slightly different from the case of the STMC. The spin-orbit-coupled equations possess a particular spin rotating symmetry,
\begin{equation}
\mathcal{R}_\text{SOC}=  \mathcal{P} e^{i\pi F_x}=\mathcal{P} \begin{pmatrix} 0 & 0&-1 \\ 0&-1&0 \\-1 &0&0 \end{pmatrix},
\end{equation}
which must be the joint of spin rotation and parity. The equations also have the symmetry $\mathcal{P}\mathcal{K} $ which is same as the spin-tensor-momentum-coupled case, so the parity of real and imaginary parts of spin-orbit-coupled solitons are even and odd respectively, which can be confirmed from Fig.~\ref{Soliton}. The eigen-equation of $\mathcal{R}_\text{SOC}$ is $\mathcal{R}_\text{SOC}\Psi(x)=\pm \Psi(x) $, taking into account the parity of real and imaginary parts of wave functions, solitons choose the eigen state with $-1$ eigenvalue, if they choose the state with $+1$ eigenvalue, then $\langle F_x \rangle =0$, the Rabi coupling energy can not be minimized. With $-1$ eigenvalue, the symmetry $\mathcal{R}_\text{SOC}$ requires that $\Psi_1(x)=\Psi_3(-x)$ and $\Psi_2(x)=\Psi_2(-x)$. Finally, because of the parity from  $\mathcal{P}\mathcal{K} $, the real parts of $\Psi_1(x)$ and $\Psi_3(x)$ become equal and the imaginary parts of $\Psi_1(x)$ and $\Psi_3(x)$ have opposite signs, while the imaginary part of $\Psi_2(x)$ must disappear.

\begin{figure*}[t]
\includegraphics[width=0.45\textwidth]{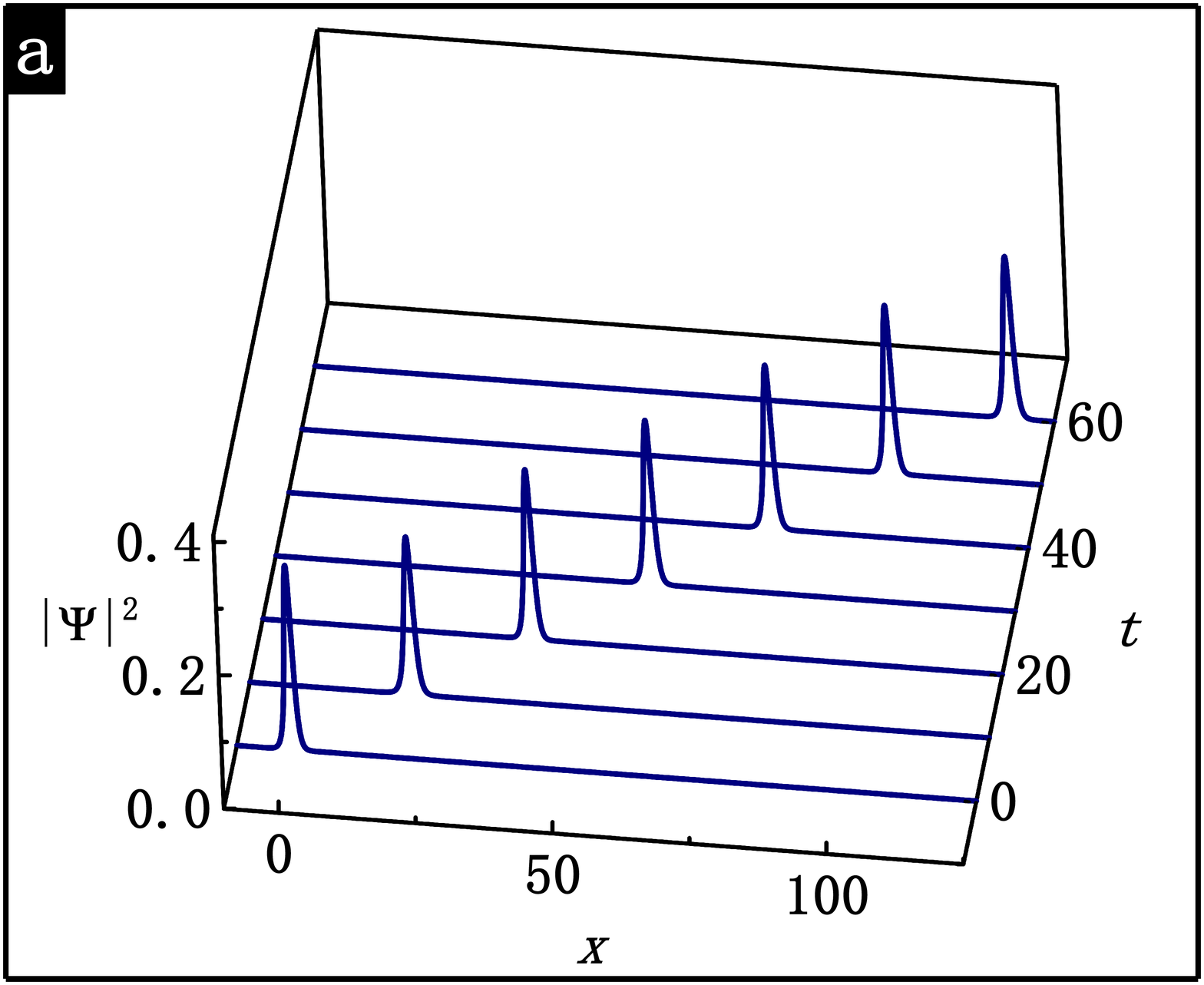}
\includegraphics[width=0.45\textwidth]{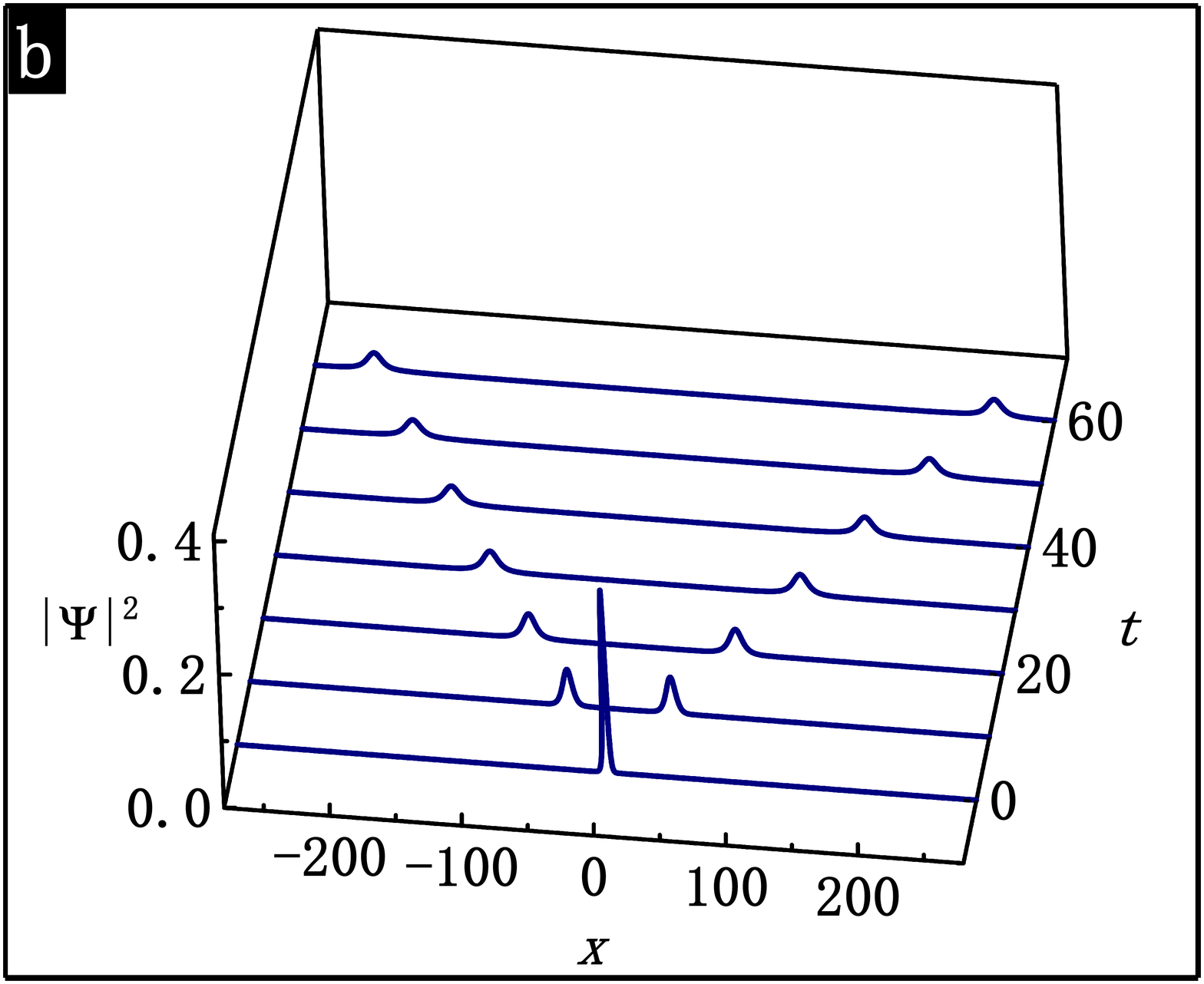}
\caption{The time evolution of an initial spin-tensor-momentum-coupled (a) and spin-orbit-coupled (b) solitons after switching off the spin-tensor-momentum coupling and spin-orbit coupling respectively, $|\Psi|^2= |\Psi_1|^2+|\Psi_2|^2+|\Psi_3|^2$. The parameters are $\Delta=-1, \Omega=0.5$ and $g_0=-2$. }
\label{Evolution}
\end{figure*}

The above symmetry analysis provides a deep insight into the understanding of solitons. So, we
are motivated to apply a variational function to stimulate corresponding solitons as follows. For the spin-tensor-momentum-coupled soliton, the variational wave function is,
\begin{equation}
\Psi_\text{STMC}=\begin{pmatrix} A[\cos(k_1x)+i\rho_0 \sin(k_1x) ] \\
    B[\cos(k_2x)+i\rho_1 \sin(k_2x) ] \\
    A[\cos(k_1x)+i\rho_0 \sin(k_1x) ] \end{pmatrix}
\text{sech}(\sigma x).
\end{equation}
This trial wave function completely satisfies the symmetries of $\mathcal{R}_\text{STMC}$ and $\mathcal{O}_\text{STMC}$. Variational parameters $A,B,k_1,k_2,\rho_0,\rho_1$ and $\sigma$ would be determined by the minimization of the total energy $E_\text{STMC}=\int dx (E_0 +\bar{E}_\text{STMC})$, with the energy density,
\begin{align}
E_0=
&|\partial_{x}\Psi_{1}|^{2}+|\partial_{x}\Psi_{2}|^{2}+|\partial_{x}\Psi_{3}|^{2}
 +(\Delta+4)( |\Psi_{1}|^{2} \notag \\
&  +|\Psi_{3}|^{2} )    +\Omega\left(\Psi_{1} \Psi_{2}^{*}+\Psi_{1}^{*} \Psi_{2}+\Psi_{2} \Psi_{3}^{*}+\Psi_{2}^{*} \Psi_{3}\right)   \notag \\
& +\frac{g_0}{2}\left(|\Psi_{1}|^{2}+|\Psi_{2}|^{2}+|\Psi_{3}|^{2}\right)^{2},
\end{align}
and
\begin{equation}
\bar{E}_\text{STMC} =4 i\left(\Psi_{1}^{*} \partial_{x} \Psi_{1}+\Psi_{3}^{*} \partial_{x} \Psi_{3}\right).
\end{equation}
Considering the symmetries of $\mathcal{R}_\text{SOC}$ and $\mathcal{P}\mathcal{K} $, the variational wave function for a spin-orbit-coupled soliton might be,
\begin{equation}
\Psi_\text{SOC}=\begin{pmatrix} A[\cos(k_1x)+i\rho_0 \sin(k_1x) ] \\
    B \cos(k_2x) \\
    A[\cos(k_1x)-i\rho_0 \sin(k_1x) ] \end{pmatrix}
\text{sech}(\sigma x).
\end{equation}
All unknown quantities appearing in above function should be determined by the minimization of  the energy $E_\text{SOC}=\int dx (E_0 +\bar{E}_\text{SOC})$, here the spin-orbit-coupled energy density is,
\begin{equation}
\bar{E}_\text{SOC} =4 i\left(\Psi_{1}^{*} \partial_{x} \Psi_{1}-\Psi_{3}^{*} \partial_{x} \Psi_{3}\right).
\end{equation}
The results from variational approximation approch  for both spin-tensor-momentum-coupled and spin-orbit-coupled solitons are shown by dot-lines in Fig.~\ref{Soliton}. Obviously, the variational wave functions are consistent with the results from the imaginary-time evolution method, as discussed before.

We characterize the properties of bright solitons by the variational wave functions. The features are identified by the dependence of $k_1,k_2,\sigma$ and the total energy $E_\text{STMC}$ and $E_\text{SOC}$ on the variables of $\Delta$ and $\Omega$. The results are described in Fig.~\ref{Feature}. The magnitudes of $k_1$ and $k_2$ are relevant to the number of nodes in soliton profiles. The larger $k_1$ and $k_2$ induce more oscillations in real and imaginary parts of soliton wave functions (see Fig.~\ref{Soliton}). This type of oscillation is  the exotic properties of STMC ($4i\partial_{x}F_{z}^{2}$) and SOC ($4i\partial_{x}F_{z}$). Because of the competition between $4i\partial_{x}F_{z}^{2}$  ($4i\partial_{x}F_{z}$ ) and $(\Delta+4) F_z^2$ or $\sqrt{2} \Omega F_x$, large $\Delta$ and $\Omega$ suppress the effect of the STMC and SOC, thus reducing the oscillation nodes. As a result, $k_1$ or $k_2$ decreases with increasing $\Delta$ or $\Omega$. This somehow explains the tendency of lines in Fig.~\ref{Feature}(a1-d2). Besides, the modification of $k_1$ and $k_2$, the Rabi coupling $\sqrt{2} \Omega F_x$ also makes soliton wave packets more spatially localized to reduce oscillations. Finally, as shown in Fig.~\ref{Feature}(b3,d3), $\sigma$ increases when
increasing $\Omega$. However, the dependence of $\sigma$ on $\Delta$ is not monotonous at all (see Fig.~\ref{Feature}(a3,c3)), resulting in two obvious slopes in the total energy as a function of $\Delta$ in Fig.~\ref{Feature}(a4,c4). Fig.~\ref{Feature}(b4,d4) demonstrates that $ \Omega$ always reduces the total energy, due to the fact that the Rabi coupling energy is proportional to $\langle F_x \rangle$, satisfying $\langle F_x \rangle<0$.

\begin{figure*}[t]
\includegraphics[width=0.7\textwidth]{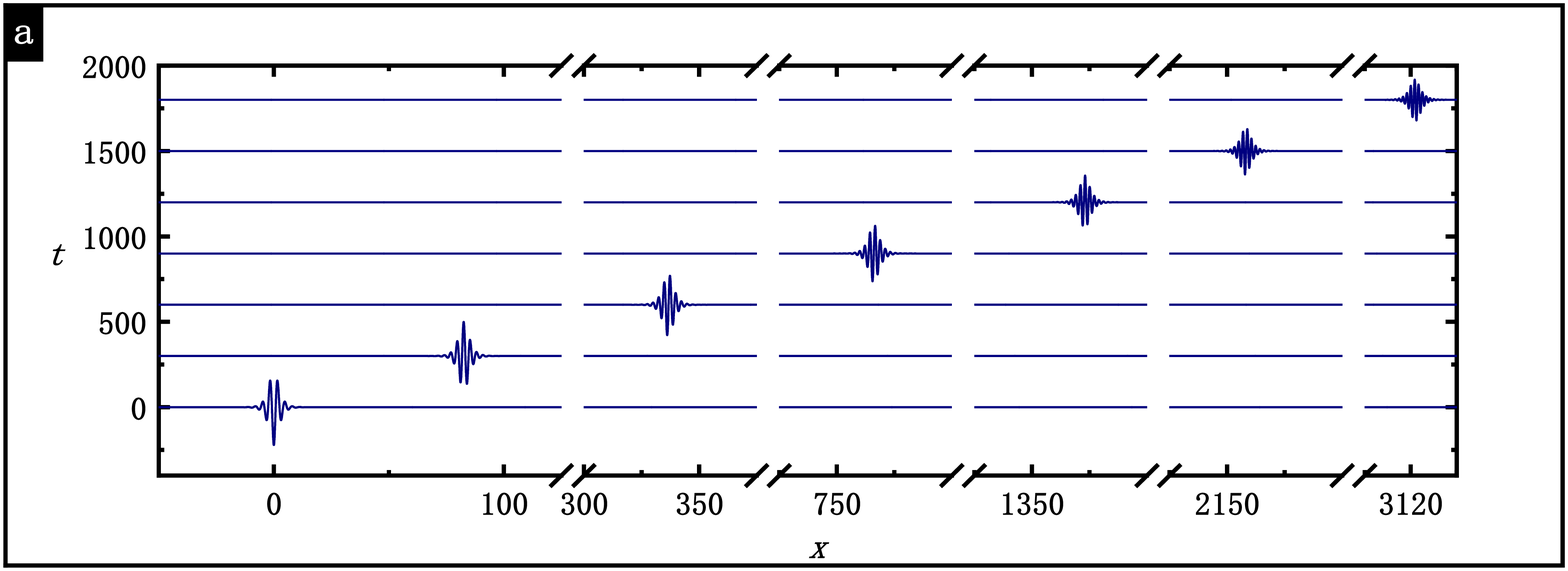}
\includegraphics[width=0.7\textwidth]{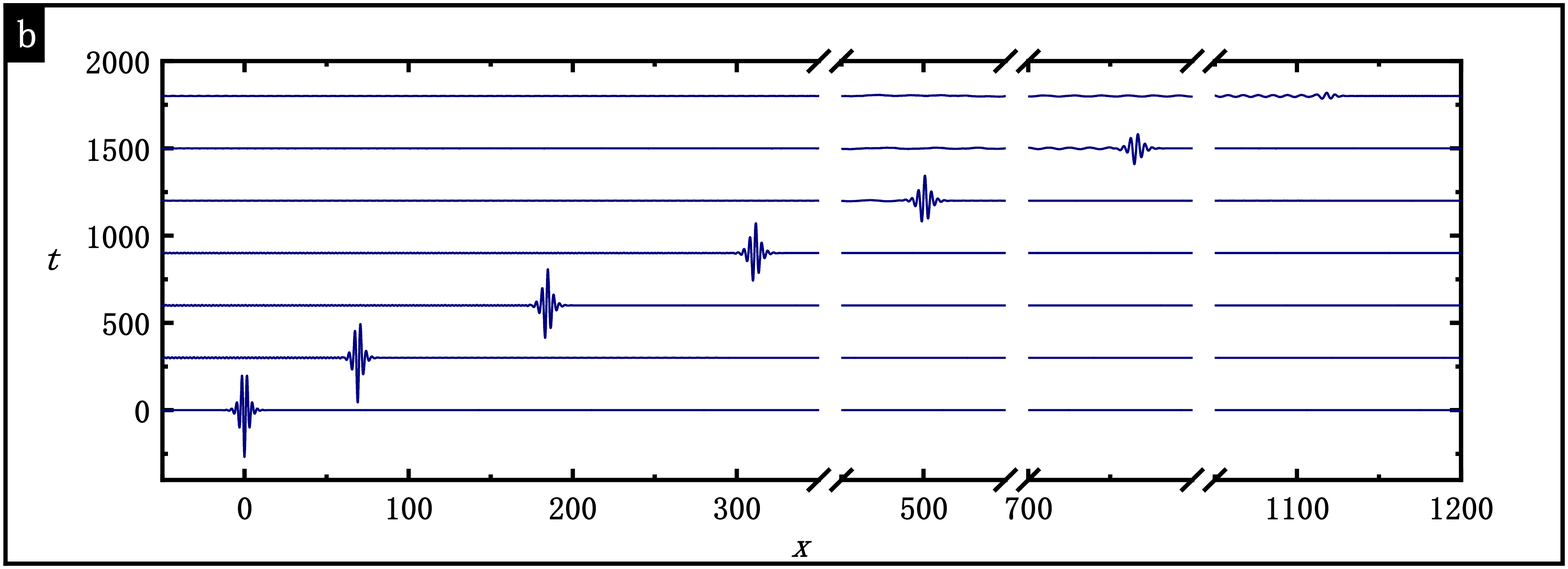}
\caption{The time evolution of an initial spin-tensor-momentum-coupled (a) and spin-orbit-coupled (b) initiated by a constant weak acceleration force which is implemented by adding a linear potential $-0.001x$ into GP equations in Eq.~(\ref{GPSTMC}) and Eq.~(\ref{GPSOC}).The parameters are $\Delta=-1, \Omega=0.5$ and $g_0=-2$.}
\label{Acceleration}
\end{figure*}

Next, we turn to the dynamics of  spin-tensor-momentum-coupled and spin-orbit-coupled solitons. Two different kinds of dynamics
are presented as follows. First of all,
the quench dynamics is shown in Fig.~\ref{Evolution},
where the initial soliton states are evolved
after switching off the STMC or SOC,
by solving the real time evolution of Eqs.~(\ref{GPSTMC}) or Eq.~(\ref{GPSOC}) but without the STMC ($4i\partial_{x}F_{z}^{2}$ term) or the SOC ($4i\partial_{x}F_{z}$ term).
Fig.~\ref{Evolution}(a) and (b) correspond to the spin-tensor-momentum-coupled and the spin-orbit-coupled solitons, respectively.
After switching off the STMC or SOC, solitons are not stationary. This provides clear evidence that the solitons are intrinsically supported by the STMC or SOC.
Interestingly, the time evolution of the spin-tensor-momentum-coupled and spin-orbit-coupled solitons are much different. The spin-tensor-momentum-coupled soliton moves along one direction, while the spin-orbit-coupled soliton splits into two parts with opposite velocities. This is because that the initial soliton satisfies $k_1=k_2,\rho_0=\rho_1=1$. Therefore, the spin-tensor-momentum-coupled soliton is the spatial confinement of a plane wave, after tuning off the STMC, it moves in the direction of the plane wave. While, the spin-orbit-coupled solion includes two plane-wave modes due to the component $\Psi_2 \propto \cos(k_2x)=(e^{ik_2x}+ e^{-ik_2x})/2$. The Rabi coupling transfers these two plane-wave modes into other components, leading to the splitting of  two branches during the evolution.

Secondly, we shall explore the acceleration of bright solitons. We add a constant weak force to accelerate the initially prepared soliton. The slow adiabatic acceleration connects moving bright solitons to stationary bright solitons~\cite{Xuyong1}. Due to the lack of Galilean invariance in spin-tensor-momentum-coupled and spin-orbit-coupled systems, the profiles of moving solitons becomes different from these of stationary solitons, therefore, they are changed during the acceleration, as illustrated in Fig.~\ref{Acceleration}. The change of the spin-orbit-coupled soliton is more pronounced than that of the spin-tensor-momentum-coupled soliton (see Fig.~\ref{Acceleration}). We provide a simple insight into the understanding of such difference. The moving bright soliton solutions should be
\begin{equation}
\Psi(x,t)= \Phi_v(x-2vt,t)e^{ivx-iv^2t},
\end{equation}
with $v$ being moving velocity. Substituting this ansatz into GP equations in Eq.~(\ref{GPSTMC}) and Eq.~(\ref{GPSOC}), the resulted equations for $\Phi_v(x-2vt,t)$ are different from the original ones by additional appearing of $-4vF_z^2$ and $-4vF_z$ respectively for the spin-tensor-momentum-coupled and spin-orbit-coupled equations. The additional $-4vF_z^2$ does not have an effect on the symmetry $\mathcal{R}_\text{STMC}$, so the moving spin-tensor-momentum-coupled bright solitons still possess $\mathcal{R}_\text{STMC}$. In contrast,  $-4vF_z$  for the spin-orbit-coupled solitons breaks the symmetry $\mathcal{R}_\text{SOC}$. The constant acceleration force linearly increases the velocities of solitons. However, the symmetry $\mathcal{R}_\text{STMC}$ manages to protect the profiles of bright soliton,
by avoiding to dramatic change. The initial stationary spin-orbit-coupled bright soliton changes distinctly during the acceleration, since the symmetry of the stationary one is so different from that of the moving one.

\section{Conclusion}
\label{conclusion}

We systematically study bright solitons in three-component BECs with the spin-tensor-momentum coupling and spin-orbit coupling, motivated by
the rapid development of the research field of spin-orbit-coupled ultracold atomic gases and by the recent proposal to realize the spin-tensor-momentum-coupled BEC. The slight difference between the STMC and SOC leads to various symmetries, which gives rise to different profiles of soliton wave functions. Moreover, the dynamics of spin-tensor-momentum-coupled and spin-orbit-coupled solitons are different during the time-evolution, when they are initiated by switching off the couplings or by a constant weak acceleration force. We conclude that all different  properties comes from different symmetries.

\section*{Acknowledgement}

We sincerely acknowlege Yong Xu, Biao Wu and Ray-Kuang Lee for helpful discussions. The work supported by the NSF of China (Grant Nos. 11974235, 1174219 and 11474193), the Thousand Young Talents Program of China, SMSTC (18010500400 and 18ZR1415500), and the Eastern Scholar and Shuguang (Program No. 17SG39) Program. XC also thanks the support from Ram\'{o}n y Cajal program of the Spanish MINECO  (RYC-2017-22482).

\bibliographystyle{apsrev4-1}

\end{document}